\def\vmu{\mbox{\boldmath$\mu$}}
\def\vsig{\mbox{\boldmath$\sigma$}}
\newcommand{\gsim}{\lower.7ex\hbox{$
    \;\stackrel{\textstyle>}{\sim}\;$}}
\newcommand{\lsim}{\lower.7ex\hbox{$
    \;\stackrel{\textstyle<}{\sim}\;$}}
\def\order#1{{\cal O}\left(#1\right)}

\newcommand{\ba}{\begin{eqnarray}}
\newcommand{\ea}{\end{eqnarray}}

\documentclass[12pt]{article}
\usepackage{epsfig}

\newcommand\pubdate{May 2001}
\newcommand\hepnumber{hep-ph/0105118}
\newcommand\pubnumber{Alberta Thy 09-01}

\def\Title#1{\begin{center} {\Large\bf #1 } \end{center}}
\def\Author#1{\begin{center}{ \sc #1} \end{center}}
\def\Address#1{\begin{center}{ \it #1} \end{center}}

\newcommand\pubblock{\rightline{\begin{tabular}{l} \pubnumber\\
         \pubdate\\ \hepnumber \end{tabular}}}
\newenvironment{Abstract}{\begin{quotation}  }{\end{quotation}}
\newenvironment{Presented}{\begin{quotation} \begin{center} 
             Presented at the\end{center}
      \begin{center}\begin{large}}{\end{large}\end{center} \end{quotation}}
\def\Acknowledgments{\bigskip  \bigskip \begin{center}
          \large\bf Acknowledgments\end{center}}

\makeatletter
\def\section{\@startsection{section}{0}{\z@}{5.5ex plus .5ex minus
 1.5ex}{2.3ex plus .2ex}{\large\bf}}
\def\subsection{\@startsection{subsection}{1}{\z@}{3.5ex plus .5ex minus
 1.5ex}{1.3ex plus .2ex}{\normalsize\bf}}
\def\subsubsection{\@startsection{subsubsection}{2}{\z@}{-3.5ex plus
-1ex minus  -.2ex}{2.3ex plus .2ex}{\normalsize\sl}}

\renewcommand{\@makecaption}[2]{%
   \vskip 10pt
   \setbox\@tempboxa\hbox{\small #1: #2}
   \ifdim \wd\@tempboxa >\hsize     
       \small #1: #2\par          
     \else                        
       \hbox to\hsize{\hfil\box\@tempboxa\hfil}
   \fi}

 \def\citenum#1{{\def\@cite##1##2{##1}\cite{#1}}}
 
\newcount\@tempcntc
\def\@citex[#1]#2{\if@filesw\immediate\write\@auxout{\string\citation{#2}}\fi
  \@tempcnta\z@\@tempcntb\m@ne\def\@citea{}\@cite{\@for\@citeb:=#2\do
    {\@ifundefined
   {b@\@citeb}{\@citeo\@tempcntb\m@ne\@citea\def\@citea{,}{\bf ?}\@warning
       {Citation `\@citeb' on page \thepage \space undefined}}%
    {\setbox\z@\hbox{\global\@tempcntc0\csname b@\@citeb\endcsname\relax}%
     \ifnum\@tempcntc=\z@ \@citeo\@tempcntb\m@ne
       \@citea\def\@citea{,}\hbox{\csname b@\@citeb\endcsname}%
     \else
      \advance\@tempcntb\@ne
      \ifnum\@tempcntb=\@tempcntc
      \else\advance\@tempcntb\m@ne\@citeo
      \@tempcnta\@tempcntc\@tempcntb\@tempcntc\fi\fi}}\@citeo}{#1}}
\def\@citeo{\ifnum\@tempcnta>\@tempcntb\else\@citea\def\@citea{,}%
  \ifnum\@tempcnta=\@tempcntb\the\@tempcnta\else
 {\advance\@tempcnta\@ne\ifnum\@tempcnta=\@tempcntb \else\def\@citea{--}\fi
    \advance\@tempcnta\m@ne\the\@tempcnta\@citea\the\@tempcntb}\fi\fi}
\makeatother

\begin{document}
\begin{titlepage}
\pubblock

\vfill
\def\thefootnote{\fnsymbol{footnote}}
\Title{Radiative corrections to bound-state properties in QED}
\vfill
\Author{Andrzej Czarnecki}
\Address{Department of Physics, University of Alberta\\
Edmonton, AB, Canada T6G 2J1}

\begin{Abstract}
This talk summarizes some results on bound-state properties obtained
since the previous (1998) RADCOR meeting.  
Recent results on radiative corrections to positronium decay width and
the $g$-factor of a bound electron are described.  A new approach to
evaluating recoil corrections in systems consisting of particles with
different masses is discussed.  
\end{Abstract}
\vfill
\begin{Presented}
5th International Symposium on Radiative Corrections \\ 
(RADCOR--2000) \\[4pt]
Carmel CA, USA, 11--15 September, 2000
\end{Presented}
\vfill
\end{titlepage}
\def\thefootnote{\arabic{footnote}}
\setcounter{footnote}{0}

\setlength{\unitlength}{1mm}


\section{Introduction}

Studies of bound-state properties in Quantum Electrodynamics (QED) are
important on their own for a variety of practical applications, but
also stimulate development of theoretical tools useful in many other
areas of physics.  High-precision QED calculations, including
evaluation of high orders in the perturbation theory, are necessary to
match the precision of atomic physics experiments, particularly in the
spectroscopy of simple atoms.  In those cases where theory and
experiment can both attain high accuracy, comparisons of measurements
with predictions often enable determination of various fundamental
physical constants, such as the fine structure constant, masses or
mass ratios of the electron, muon, and proton, ratios of various
magnetic moments, proton charge radius, etc.  

In other cases, when the relevant constants are known from other
sources, one can very precisely test theoretical understanding of
bound states.  An impressive example, discussed at the previous RADCOR
meeting \cite{Czarnecki:1998gw}, is the hyperfine splitting in the
positronium ground state.  There, the two-loop effects change the
leading order prediction by about 12 MHz, or only 0.006\%, but this
effect is still  larger than the experimental error by more than an
order of magnitude!  Rarely is computing of high-order loop effects
more rewarding than it is in the spectroscopy of simple atoms like the
positronium or muonium.  

Such high-precision comparisons of theory and experiment are possible
in simple atoms because on the one hand, measurements can be made
very accurately with the modern spectroscopic methods, and on
the other there are hardly any principal limitations of the theory.
Since electrons are much lighter than any hadrons, the computations
are not hindered by non-perturbative QCD uncertainties.  Spectra and
lifetimes of simple atoms can, in principle, be evaluated with any
accuracy required by current experiments within pure QED.  

Theoretical tools developed in this way, such as the computational
techniques for the Feynman integrals or the  machinery of
non-relativistic effective theories \cite{Caswell:1986ui}, can
subsequently be applied to solve problems in other areas of physics,
such as hadronic properties or thermal field theory.  

In this talk I would like to present some results obtained in the last
couple of years since the previous RADCOR meeting in Barcelona.  Those
results include the positronium lifetime, bound electron gyromagnetic
factor, and a new approach to computing properties of bound states
consisting of particles with widely different masses.


\section{Positronium decay}
Lifetimes of the singlet and triplet positronium ground-states, p-Ps
and o-Ps, can be measured with high precision \cite{CzKar}.  For
several years there was a very significant discrepancy between the
theoretical predictions for the o-Ps lifetime and the experimental
results.  More recently, studies performed at the University of Tokyo
gave results consistent with the theory.  The most accurate
experimental results are summarized in Table \ref{Tortho}.  Recently,
the results obtained in gases were reexamined and new corrections were
taken into account.  The preliminary updated central value for the
o-Ps decay rate measured in this method is approximately 
7.047/$\mu$s \cite{ContiHydr}, in very good agreement with the vacuum
measurement, but still significantly higher than the QED prediction.

\begin{table}[htb]
\caption{Recent experimental results for the o-Ps lifetime.  ``Method''
in the second column refers to the medium in which the o-Ps decays.  The
last column shows the value of the two-loop coefficient $B_o$,
necessary to bring the theoretical prediction
(\protect\ref{eq:otheor}) into agreement with the
given experimental value.  The last line gives the present theoretical
prediction.}
\label{Tortho}
\begin{center}
\begin{tabular}{l @{\hspace*{10mm}}c @{\hspace*{10mm}}l @{\hspace*{10mm}}r}
\hline \hline
&&&\\
Reference & Method & $\Gamma(\mbox{o-Ps})~[\mu s^{-1}]$ & $B_o$ \\
&&&\\
\hline
&&&\\
Ann Arbor \protect{\cite{Westbrook89}}&Gas&7.0514(14)&338(36)\\
&&&\\
Ann Arbor \protect{\cite{Nico:1990gi}} & Vacuum&7.0482(16)&256(41)\\
&&&\\
Tokyo \protect{\cite{Asai:1995re}} & SiO$_2$ powder&7.0398(29)&41(74)\\
&&&\\
Tokyo (preliminary) \protect{\cite{Asai:2001}} & SiO$_2$ powder&
                                                    7.0398(15)&41(37)\\
&&&\\
\hline
\multicolumn{2}{c}{}&&\\
\multicolumn{2}{c}{QED prediction 
              \protect\cite{Adkins:2000fg}} & 7.0399 & 44.87(26)\\
\multicolumn{2}{c}{}&&\\[-1mm]
\hline
\hline
\end{tabular}
\vspace*{2mm}
\end{center}
\end{table}

The gas and vacuum results led to the suspicion that the two-loop QED
corrections, which were not fully known, may be very large.  The
coefficient of $(\alpha/\pi)^2$ in the correction relative to the
lowest-order decay rate is denoted by $B_o$.  Its value, necessary to
reconcile a given experimental result with the QED prediction, is
given in the last column of Table \ref{Tortho}.

If such unusually enhanced effects existed, one would expect them to
modify the p-Ps decay rate as well.  Since p-Ps decay is much simpler
than that of o-Ps, we undertook a full two-loop QED study of this
process \cite{Czarnecki:1999gv,Czarnecki:1999ci}.  

The parapositronium decay rate into two photons agrees with 
predictions and had previously attracted less theoretical
attention.  However, it is also measured sufficiently precisely
\cite{AlRam94},
\begin{equation}
\Gamma_{\rm  p-Ps}^{\rm  exp} ({\rm  gas})
= 7990.9(1.7)~{\rm \mu s}^{-1},
\label{eq:pPsEx}
\end{equation}
to warrant a calculation of ${\cal O}(\alpha_s^2)$ corrections.
The prediction for this decay width can be parameterized as
\ba
\Gamma_{\rm p-Ps} &=& {m\alpha^5\over 2}
\left[ 1+A_p{\alpha\over \pi}
+2\alpha^2\ln{1\over \alpha}
+B_p\left( {\alpha\over \pi} \right)^2
-{3\alpha^3\over 2\pi}\ln^2{1\over \alpha} \right.
\nonumber
\\
&& \left.
+{\alpha^3\over\pi}\ln\alpha\left(10\ln 2-{367\over 90}-2A_p\right)
+\ldots\right], \qquad
A_p={\pi^2\over 4}-5.
\ea
Our aim was the evaluation of
the second order non-logarithmic correction $B_p$.
It receives contributions from both  soft and hard scales,
$
 B_p = B_p^{\rm soft} + B_p^{\rm hard}.
$
We found
\begin{eqnarray}
&& B_p^{\rm  soft}  + 2 \pi^2  \ln \frac {1}{\alpha} = 
 \frac{\pi^2}{2 \epsilon}+2 \pi^2  
\ln  \frac {1}{m \alpha} + \frac {107 \pi^2}{24},
\nonumber \\
&& B_p^{\rm  hard} = -\frac {\pi^2}{2 \epsilon} 
+ 2 \pi^2  \ln(m) - 40.46(30) + {1\over 4}A_p^2,
\end{eqnarray}
and the final result is $ B_p=5.1(3)$.
There is a significant cancelation 
between soft and hard pieces and the final result is almost eight times 
smaller  than the magnitude of the finite constant in the 
hard scale contribution computed in dimensional 
regularization.

Using the above result for $B_p$ we arrive at the following result for the 
decay rate:
$$
\Gamma_{ \rm  p-Ps}^{\rm  theory} = 
7989.64(2){{\rm  \mu s}}^{-1},
$$
which agrees very well with the measured value, Eq.~(\ref{eq:pPsEx}).

Most recently, the two-loop corrections were also computed for the
o-Ps decay,
\begin{eqnarray}
\Gamma_{\rm o-Ps} &=& m\alpha^6 {2(\pi^2-9)\over 9\pi}
\left[ 1-A_o{\alpha\over \pi}
-{\alpha^2\over 3}\ln{1\over \alpha}
+B_o\left( {\alpha\over \pi} \right)^2
-{3\alpha^3\over 2\pi}\ln^2{1\over \alpha}\right.
\nonumber
\\
&& \left.
-{\alpha^3\over \pi}\ln{1\over \alpha}
      \left(8\ln 2-{229\over 30} +{A_o\over 3}\right)
+ \ldots\right]
\nonumber
\\
&=& 7.03994(1)/\mu s, \qquad A_o = 10.286606(10).
\label{eq:otheor}
\end{eqnarray}
The non-logarithmic two-loop coefficient is (with the
small light-by-light contribution shown explicitly) $B_o =
44.52(26)+0.350(4) = 44.87(26)$ \cite{Adkins:2000fg,Adkins:2001}. 
For both o-Ps and p-Ps, the leading logs in the order $\alpha^3$ were
found in \cite{DL} and the next-to-leading
logarithms were computed only recently
\cite{Melnikov:2000fi,Hill:2000qi,Kniehl:2000dh}.

Can we claim that the positronium lifetime puzzle has been solved?  It
would certainly be very valuable to have another measurement of the
o-Ps lifetime, especially that the powder measurement is somewhat
controversial.  An independent calculation of the theoretical
prediction would also be useful, although it is unlikely that any
unusually large effects are there to be uncovered.

\section{Expansion of bound-state energies in the
constituent mass ratio}

A new approach to computing energy levels of a
non-relativistic bound-state of two constituents with masses $M$ and
$m$, by a systematic expansion in powers of $m/M$, was described in
\cite{Czarnecki:2000fv}.

Simple atoms of experimental interest often consist of two particles
(constituents), widely separated in mass.  An extreme example is the
hydrogen, where the ratio of the proton and electron masses is of the
order of 2000.  Smaller ratios characterize muonium, muonic hydrogen,
and exotic hadronic atoms.  The goal of \cite{Czarnecki:2000fv} was to
find a practical algorithm which would allow evaluation of the
bound-state energy levels (in a given order of perturbation theory in
$\alpha$ and $Z\alpha$) as an expansion in powers and logarithms of
$m/M$ with an arbitrary precision.

That algorithm is useful in finding the so-called ``hard-scale
corrections''.  In the language of an effective field theory, this
corresponds to determining the Wilson coefficients of the
short-distance operators, generated by virtual momenta much larger
than the characteristic momenta of the atomic constituents.  In
practical terms, what is needed is the scattering amplitude of the two
particles, with masses $m$ and $M$, at the threshold, that is when the
particles have vanishing velocities.  The relevant Feynman diagrams
depend on only the two mass scales $m$ and $M$ (since the external
spatial momenta vanish).  Since, however, at the two-loop level, which
is of interest for the current theoretical studies, such integrals
cannot in general be evaluated exactly, we need a method of expanding
them in powers and logs of $m/M$.  

As is already well known in the theory of asymptotic expansions (for
reviews and further references see
e.g. \cite{Chetyrkin91,Tkachev:1994gz,Smirnov:1995tg}), such expansion
consists first of all in expanding the Feynman integrand in the small
parameter $m$.  In general, this gives rise to non-integrable
singularities at small values of some momenta for which, before the
expansion, the mass $m$ provided a regulator.  The difficulty in
constructing a correct algorithm is to find the necessary counterterms
and to evaluate the resulting new integrals.  

\begin{figure}[htb]
\vspace*{3mm}

\hspace*{50mm}
\psfig{figure=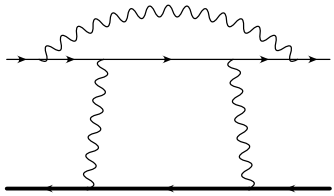,width=43mm}

\caption{An example of a forward-scattering radiative-recoil
diagram. The bold line represents the heavy constituent of the
bound-state (e.g. proton if we consider hydrogen) and the thin line
--- the light one (an electron).}
\label{fig1}
\end{figure}

Such a procedure was carried out in \cite{Czarnecki:2000fv} with the
example of radiative-recoil corrections, such as the diagram shown in
Fig.~\ref{fig1}.

\begin{figure}[htb]
\vspace*{3mm}

\hspace*{21mm}
\begin{tabular}{cc}
\psfig{figure=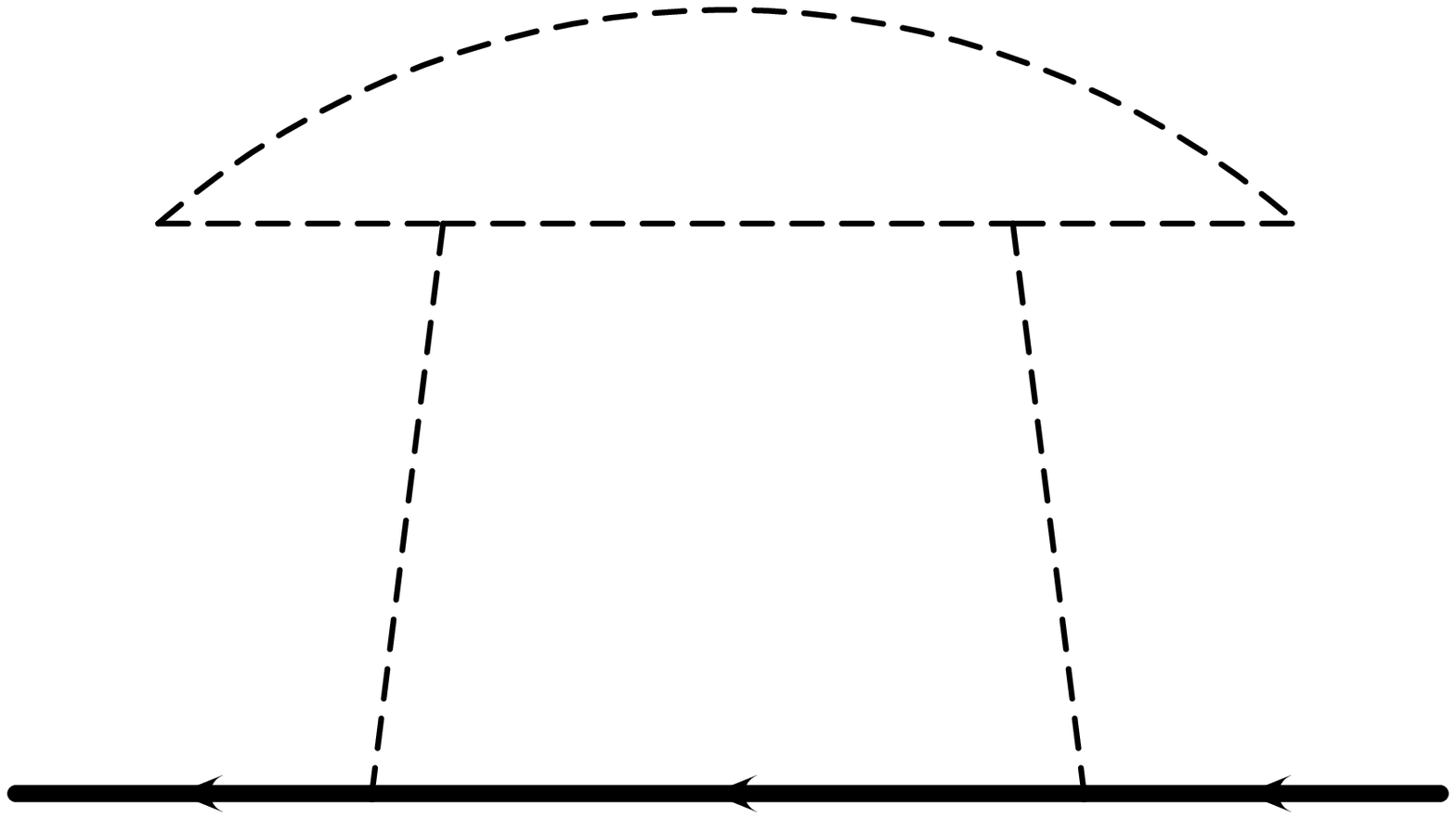,width=44mm}
&\hspace*{8mm}
\psfig{figure=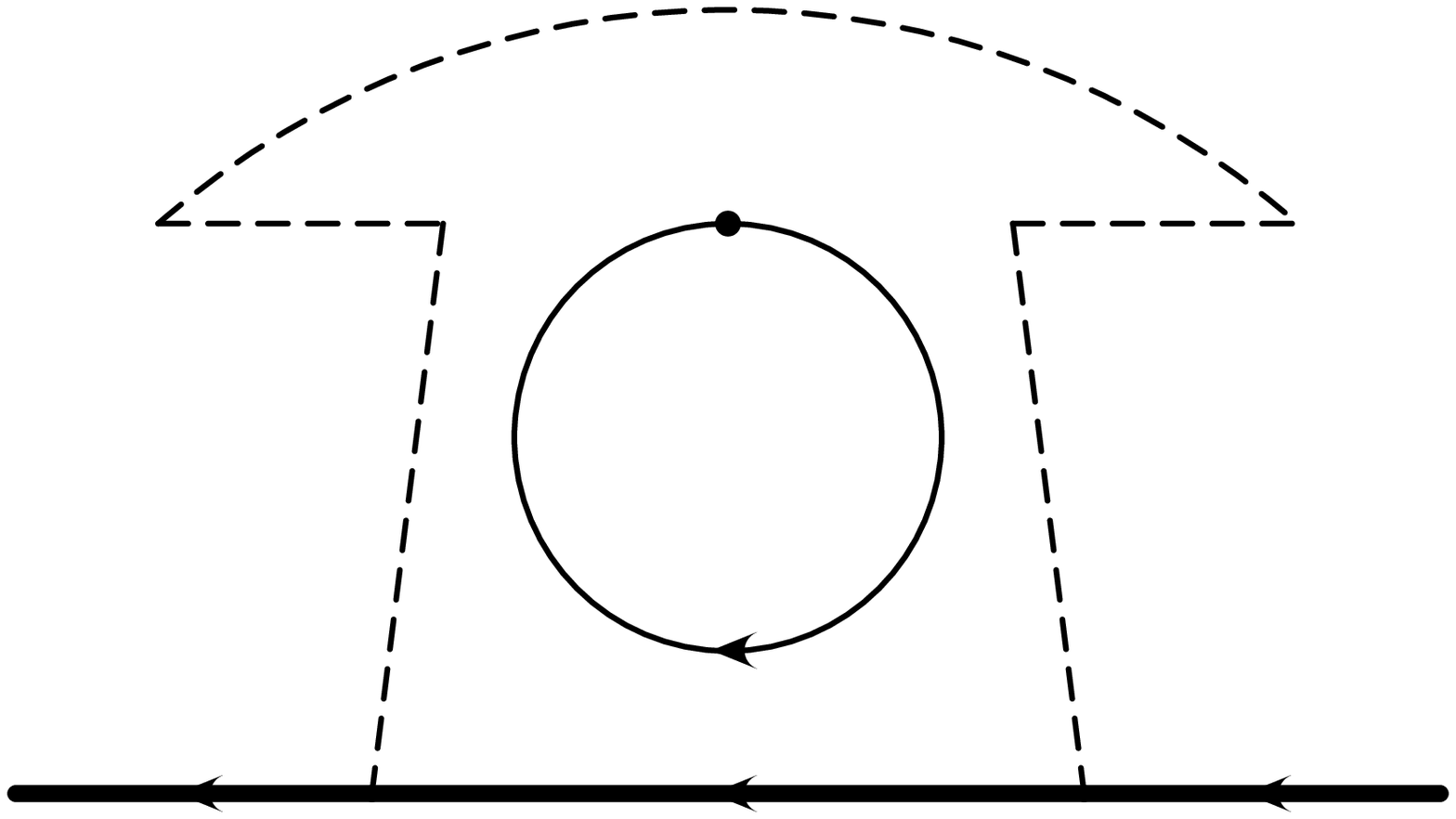,width=44mm}
\\[2mm]
(a) &\hspace*{8mm} (b) 
\\[6mm]
\psfig{figure=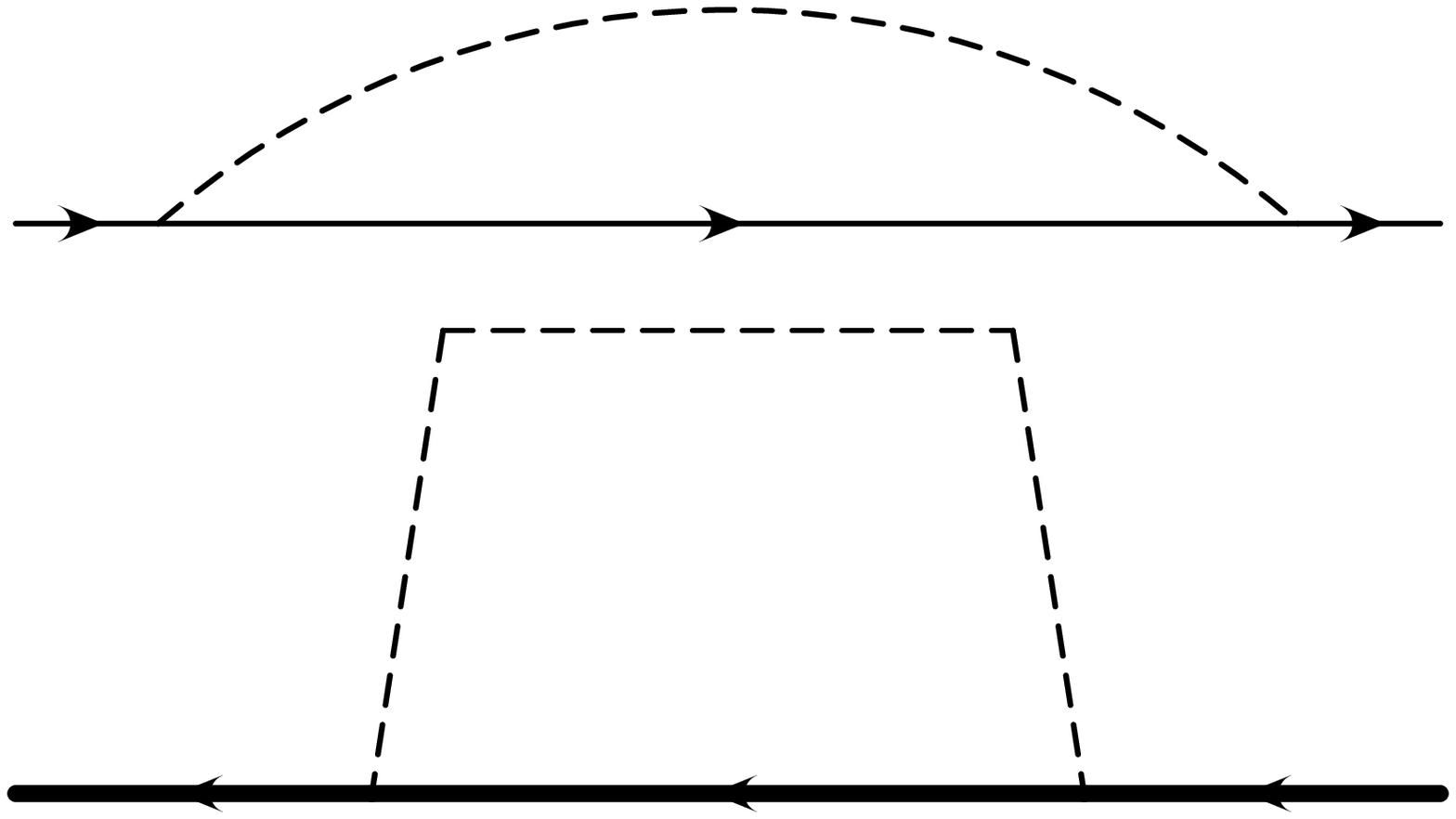,width=44mm}
&\hspace*{8mm}
\psfig{figure=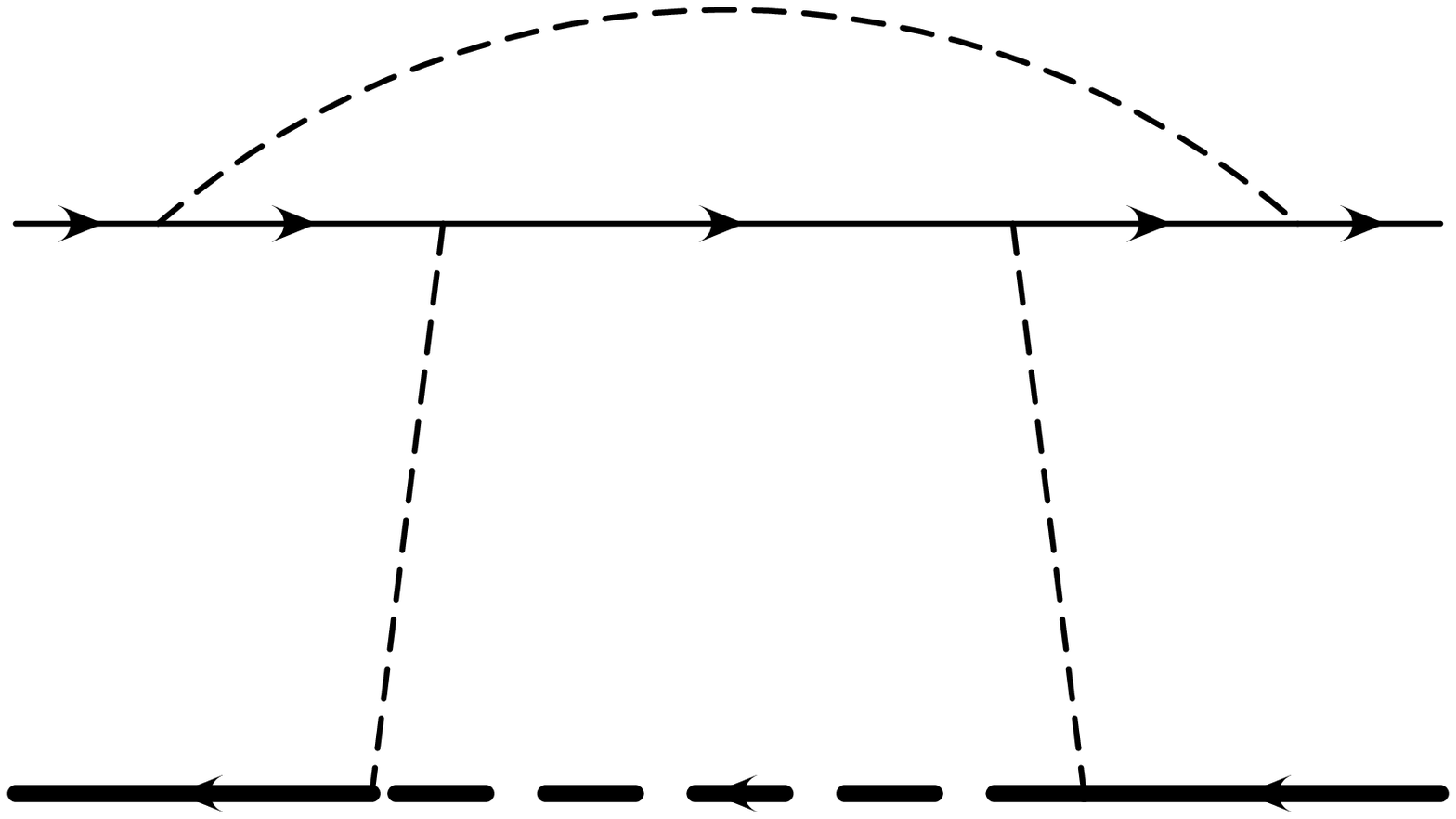,width=44mm}
\\[2mm]
(c) &\hspace*{8mm} (d) 
\end{tabular}

\caption{Elements of the expansion of the diagram in
Fig.~\protect\ref{fig1} in powers and logs of $m/M$ (see
text).  Thick and thin solid lines denote propagators with $M$ and
$m$, respectively.  Thin dashed lines are massless, and the thick
dashed line denotes a static (``eikonal'') propagator.} 
\label{fig2}
\end{figure}

In the language of characteristic scales of the momenta in the
counterterm integrals, there are 3 ``regions''.  Fig.~\ref{fig2}
depicts the Taylor expansion in $m$ (\ref{fig2}(a)) and the 3
counterterms.  In the diagram \ref{fig2}(b), the momentum flowing
through one of the light constituent lines is of the order of $m$,
while the second loop momentum is $\order{M}$ (so that the propagators
along its flow can be Taylor-expanded in $m$ and in the corresponding
small momentum).  In \ref{fig2}(c) we have the same sizes of the loop
momenta, but now the large momentum takes a different route to flow
through the upper part of the diagram.  The last contribution,
\ref{fig2}(d), arises when both loop momenta are $\order{m}$.  In this
case, the heavy propagator $1/(k^2+2k.P)$ can be expanded in $k^2$,
and becomes an ``eikonal'' \cite{CzSmir} propagator $1/2k.P$.

This last contribution was least easy to evaluate and a
procedure for computing the resulting integrals was described in
\cite{Czarnecki:2000fv}.  It is interesting that the same integrals
arise in very different problems, for example in certain corrections
to the radiative quark decays \cite{BCMU}.

\section{Anomalous magnetic moment of a bound electron}
In \cite{Czarnecki:2000uu} the binding corrections to the gyromagnetic
factor $g_{\rm e}$ of an electron in hydrogen-like ions were studied.
The interaction of an electron with an external magnetic field
{\boldmath$B$} is described by the potential $V = -\vmu \cdot
\mbox{\boldmath$B$}$.  The electron magnetic moment $\vmu$ is
given by $\vmu = g_{\rm e}\, {e\over 2m} \mbox{\boldmath$s$}$, with
$m$, $\mbox{\boldmath$s$} = \vsig/2$, and $g_{\rm e}$ denoting the
mass, spin and the so-called gyromagnetic or Land\'e factor of the
electron.

If the electron is bound in a ground state of a hydrogen-like ion,
$g_{\rm e}$ becomes a function of the nuclear charge $Z$ and its
measurements \cite{hermanspahn96,Quint95,Werth95} provide a sensitive
test of the bound-state theory based on the QED.  The theoretical
prediction can be cast in the following form \cite{Mohr99}
\ba
g_{\rm e}\,(Z) = g_{\rm D} + \Delta g_{\rm rec} + \Delta g_{\rm rad}.
\ea
The first term corresponds to the lowest order expansion in
$\alpha/\pi$ and has been calculated to all orders in $Z\alpha$
\cite{Breit28},
\ba
g_{\rm D} = {2\over 3}\left[ 1 + 2\sqrt{ 1-(Z\alpha)^2 }\right].
\label{eq:Breit}
\ea
$\Delta g_{\rm rec}$ denotes the recoil corrections
\cite{grotch70recoil}, $\Delta g_{\rm rec} = 
\order{(Z\alpha)^2 {m\over m_N}}$, where $m_N$ is the nucleus mass.
Further references to the studies of those effects can be found in
\cite{Mohr99}. 

Our main interest were the radiative corrections.
They can be presented as an expansion in two parameters, $Z\alpha$ and
$\alpha/\pi$,
\ba
{ \Delta g_{\rm rad} \over 2}  =  
   C_{\rm e}^{(2)} (Z\alpha) \left( {\alpha\over \pi}\right)
 + C_{\rm e}^{(4)} (Z\alpha) \left( {\alpha\over \pi}\right)^2 
 + \ldots
\label{eq:series}
\ea
Powers of $\alpha/\pi$ correspond to electron--electron interactions,
while $Z\alpha$ governs binding effects due to electron interactions
with the nucleus.  
The first coefficient function in (\ref{eq:series}), $C_{\rm e}^{(2)}
(Z\alpha)$, has been computed numerically to all orders in  $Z\alpha$
\cite{blundell97,Persson97}.  
Its first two terms in the $Z\alpha$ expansion are
also known analytically \cite{sch48,grotch70}
\ba
 C_{\rm e}^{(2)} (Z\alpha) = {1\over 2}\left[1 + {1\over 6}(Z\alpha)^2 +
 \order{(Z\alpha)^4}\right].
\label{eq:Grotch}
\ea
The main theoretical uncertainty for $g_{\rm e}$ in light ions is, at
present, connected with the unknown coefficient $C'$ in
the next coefficient function,
\ba
 C_{\rm e}^{(4)} (Z\alpha) &=&  C_{\rm e}^{(4)} (0)\left[1 + C'\cdot
 (Z\alpha)^2 + 
 \order{(Z\alpha)^4}\right], 
\nonumber \\
 C_{\rm e}^{(4)} (0) &=& -0.328\,478\,444\,00\ldots \qquad
 \cite{som57,pet57a,Mohr99}.
\label{eq:cprime}
\ea

At present, the most accurate experimental value of the bound electron
gyromagnetic factor has been obtained
\cite{hermanspahn00,Haeffner} 
with a hydrogen-like carbon ion $^{12}$C$^{5+}$ ($Z=6$),
\ba
g_{\rm e}(Z=6;{\rm exp}) = 2.001\,041\,596(5).
\label{eq:exp}
\ea
  The theoretical prediction is \cite{Beier}
\ba
g_{\rm e}(Z=6;{\rm theory}) = 2.001\, 041\, 591(7)
\label{eq:theoryC}
\ea
where 70\% of the error is caused by the unknown coefficient $C'$ of
the $\left({\alpha\over\pi}\right)^2 (Z\alpha)^2$ effects in
(\ref{eq:cprime}) (for carbon, higher powers of $Z\alpha$ are assumed
to be negligible).

The purpose of our paper \cite{Czarnecki:2000uu} was 
to demonstrate that $C'={1/ 6}$, in
analogy to the corresponding coefficient in the lower order in
$\alpha/\pi$.  In fact, we found that the coefficient of
$(Z\alpha)^2$ is the same in all coefficient functions
$C^{(2n)}_e(Z\alpha)$, so that the theoretical prediction for 
$\Delta g_{\rm rad}$ accurate up to $(Z\alpha)^2$ and exact in 
$\alpha/\pi$ reads
\ba
\Delta g_{\rm rad} = 
 (g_{\rm free} -2)\cdot
\left[1+ {(Z\alpha)^2\over6}\right],
\label{eq11}
\ea
where $g_{\rm free}$ is the gyromagnetic factor of a free electron,
presently known to $\order{(\alpha/\pi)^4}$ \cite{Hughes:1999fp} (the
same result had been obtained in a different 
way in \cite{EG}).
With this result, the theoretical uncertainty in (\ref{eq:theoryC}) is
reduced from $7 \cdot 10^{-9}$ to about $2\cdot 10^{-9}$.

\section{Conclusions}

Three examples of problems in the bound-state theory, solved since the
last RADCOR meeting, were summarized in this talk.  Of course, other
groups have made important progress in other aspects of bound-state
physics, which was not reported here.  I would like to mention two
examples.  
Important new corrections
to the hydrogen Lamb shift were found in
\cite{Melnikov:1999xp,Pachucki2000}.  
In \cite{Melnikov:1999xp}, an algorithm was constructed
to compute a  class of 3-loop propagator-type massive
integrals (the so-called master integrals had been computed in
\cite{Laporta:1996mq}).  

Another class of recently studied problems is connected with the
``velocity renormalization group'' \cite{Luke:1999kz}.  This approach
allows better understanding and at least partial resummation of large
logarithms $\ln{1\over \alpha}$ arising in the non-relativistic
bound-state  calculations.  A review of the recent progress in this
area can be found in \cite{Manohar:2000cg}.  

Hopefully, the recent significant progress in the theory will be
followed by new measurements.  It would be very important to
re-measure the positronium hyperfine splitting, to resolve the present
(more than 3 standard deviations) discrepancy between the old
measurements and the improved theory
\cite{Czarnecki:1998gw,Hill:2000zy,Melnikov:2000zz,Kniehl:2000cx}.    
Similarly important would be an independent
study of the ortho-positronium lifetime.  Now that the theory has in
most cases taken control of the two-loop quantum effects, it is
particularly exciting to test those predictions experimentally.

\Acknowledgments

I am grateful to Kirill Melnikov and Alexander Yelkhovsky, with whom
the results described here were obtained, for the continuing
collaboration and for helpful remarks about this paper.  This research
was supported by the Natural Sciences and Engineering Research Council
of Canada.  I thank the organizers of the RADCOR-2001, especially
Howard Haber, for partially supporting my participation.


\end{document}